\documentclass[reprint,amsmath,amssymb,aps,pre,longbibliography]{revtex4-2}

\usepackage{txfonts}
\DeclareMathAlphabet{\mathcal}{OMS}{cmsy}{m}{n}

\usepackage[utf8]{inputenc}
\usepackage[dvipsnames]{xcolor}
\usepackage[colorlinks=true,linkcolor=Blue,citecolor=Blue,urlcolor=Blue]{hyperref}
\usepackage{graphicx}
\usepackage[english]{isodate}
\cleanlookdateon

\graphicspath{{./figures/}}

\newcommand{\bnab}{\boldsymbol{\nabla}}
\newcommand{\ed}{\mathrm{d}}
\newcommand{\tr}{\operatorname{tr}}

\begin{document}

\title{Stable photon orbits in stationary axisymmetric spacetimes \\
with an electromagnetic field and a cosmological constant}

\author{Jake O.~Shipley}
\email{shipley.jake.oliver@gmail.com}
\affiliation{Government Operational Research Service, UK Civil Service, Sheffield, United Kingdom}

\date{\today}

\begin{abstract}
	Stable light rings, which are associated with spacetime instabilities, are known to exist in four-dimensional stationary axisymmetric spacetimes that solve the Einstein--Maxwell equations (so-called electrovacuum solutions, with Faraday tensor $F_{\mu \nu} \neq 0$); however, they are not permitted in pure vacuum ($F_{\mu \nu} = 0$). In this work, we extend this result to spacetimes with a non-zero cosmological constant $\Lambda$. In particular, we demonstrate that stable light rings are permitted in $\Lambda$-electrovacuum ($F_{\mu \nu} \neq 0$, $\Lambda \neq 0$), but ruled out in $\Lambda$-vacuum ($F_{\mu \nu} = 0$, $\Lambda \neq 0$). \newline
	
	\noindent DOI: \href{https://doi.org/10.1103/PhysRevD.108.084040}{10.1103/PhysRevD.108.084040} (publication)
\end{abstract}

\maketitle
\newpage

\section{Introduction \label{sec:intro}}

In general relativity, black holes and other ultra-compact objects can cause spacetime to distort in such a way that photons are forced to follow orbits. Around a rotating Kerr black hole, for example, there are two circular photon orbits -- or \emph{light rings} -- that lie in the equatorial plane. Between the inner (prograde) and outer (retrograde) light rings, the Kerr solution admits a family of spherical, i.e., constant-radius, photon orbits that are not confined to the equatorial plane, exhibiting rich latitudinal motion \cite{Teo2003}.

For the Kerr black hole, and in many other scenarios, the photon orbits are unstable: small perturbations can cause the photon to either cross the black hole's event horizon or escape to future null infinity. These unstable photon orbits play an important physical role, particularly in strong-field gravitational lensing \cite{Perlick2004}. Around black holes, for example, the unstable photon orbits determine the shape and size of the black hole shadow \cite{CunhaHerdeiro2018}.

\emph{Stable} photon orbits are also known to exist in certain contexts, e.g.~within the inner horizons of Kerr--Newman black holes and around naked singularities \cite{Liang1974, CalvaniDeFeliceNobili1980, Stuchlik1981, BalekBicakStuchlik1989, KhooOng2016}, around pairs of charged black holes in the Majumdar--Papapetrou spacetime \cite{DolanShipley2016, Shipley2019}, and in the spacetimes of horizonless ultra-compact objects \cite{CardosoEtAl2014, CunhaEtAl2016, CunhaBertiHerdeiro2017, CunhaHerdeiro2020}.

Stable photon orbits are associated with dynamical \emph{instabilities} in spacetime: it is possible for an arbitrary number of photons to pile up at a stable light ring, which would lead to a back-reaction on the spacetime \cite{CardosoEtAl2014, GhoshSarkar2021}. Indeed, in recent work, Cunha \emph{et al.} \cite{CunhaEtAl2023} performed fully non-linear evolutions of ultra-compact boson stars, confirming that stable light rings trigger spacetime instabilities. Moreover, stable photon orbits are associated with slow (logarithmic) decay of perturbations \cite{Keir2016}, a modified late-time gravitational-wave ringdown \cite{CardosoFranzinPani2016}, and distinctive signatures in gravitational lensing by black holes and ultra-compact objects \cite{Shipley2019, CunhaEtAl2016, SengoEtAl2023}.

In Ref.~\cite{DolanShipley2016}, it was demonstrated that generic stable light rings are permitted in four-dimensional stationary axisymmetric spacetimes that solve the Einstein--Maxwell equations of gravity and electromagnetism (so-called \emph{electrovacuum} solutions, with Faraday two-form $F_{\mu \nu} \neq 0$); however, it was shown that they are not permitted in pure vacuum ($F_{\mu \nu} = 0$).

It is natural to ask whether the presence of other (effective) stress--energy sources can be introduced to forbid stable light rings. In this work, we consider whether the inclusion of a non-zero cosmological constant in the Einstein--Maxwell equations is enough to prevent the existence of stable light rings, and therefore spacetime instabilities.

This paper is organised as follows. Section \ref{sec:RNadS} contains a brief review of the existence and stability of equatorial circular photon orbits in the Reissner--Nordstr\"{o}m--(anti-)de Sitter spacetime. In Sec.~\ref{sec:geometry}, we review four-dimensional solutions to the Einstein--Maxwell equations with a cosmological constant $\Lambda$, under the general assumptions of stationarity and axisymmetry. In Sec.~\ref{sec:light_rings}, we analyse light rings and their stability: first, we introduce a Hamiltonian formalism to describe null geodesics and therefore light rings (Sec.~\ref{sec:hamiltonian}); we then employ a subset of the Einstein--Maxwell field equations to classify the stability of light rings (Sec.~\ref{sec:classification}). Our main result is a generalisation of the key conclusion of Ref.~\cite{DolanShipley2016}: we demonstrate that generic stable light rings are permitted in situations with $F_{\mu \nu} \neq 0$ and $\Lambda \neq 0$; however, they are ruled out when $F_{\mu \nu} = 0$ and $\Lambda \neq 0$. We conclude with a discussion of possible extensions to this work in Sec.~\ref{sec:discussion}.

\section{Circular photon orbits in Reissner--Nordstr\"{o}m--(anti-)de Sitter spacetimes \label{sec:RNadS}}

In this section, we review the existence and stability of equatorial circular photon orbits (ECPOs) in the Reissner--Nordstr\"{o}m--(anti-)de Sitter [RN(a)dS] family of spacetimes.

The RN(a)dS geometry is a static, spherically symmetric solution to the Einstein--Maxwell equations with a non-zero cosmological constant $\Lambda$. It describes a Reissner--Nordstr\"{o}m (RN) black hole or naked singularity of mass $M$ and charge $Q$, embedded in an asymptotically de Sitter (dS) spacetime for $\Lambda > 0$, or an asymptotically anti-de Sitter (adS) spacetime for $\Lambda < 0$.

In Schwarzschild coordinates $\{ t, r, \theta, \phi \}$, the metric is given by
\begin{equation}
	\ed s^{2} = - W(r) \, \ed t^{2} + \frac{\ed r^{2}}{W(r)} + r^{2} ( \ed \theta^{2} + \sin^{2}{\theta} \, \ed \phi^{2} ) ,
\end{equation}
with
\begin{equation}
	\label{eqn:RNadS_metric_function}
	W(r) = 1 - \frac{2 M}{r} + \frac{Q^{2}}{r^{2}} - \frac{\Lambda}{3} r^{2} ,
\end{equation}
and the only non-zero component of the electromagnetic four-potential $A_{\mu}$ in the standard gauge is $A_{t} = Q/r$.

The horizons of the RN(a)dS spacetime correspond to the roots of $W(r)$. These can be found by considering the discriminant of the polynomial obtained by recasting $W(r)$ in terms of the coordinate $u = M/r$, and multiplying through by $u^{2}$. This discriminant turns out to be $\mathcal{D} ( q^{2}, \ell ) = -16 \ell \mathcal{B} ( q^{2}, \ell )$, where we have introduced the quantity
\begin{equation}
	\label{eqn:horizon_discriminant}
	\mathcal{B} ( q^{2}, \ell ) = 16 \ell^{2} q^{6} + 8 \ell q^{4} - 36 \ell q^{2} + q^{2} + 27 \ell - 1 .
\end{equation}
Here, we have defined the rescaled parameters $q = Q/M$ and $\ell = \Lambda M^{2}/3$.

\begin{figure}
	\centering
	\includegraphics[width=8.6cm]{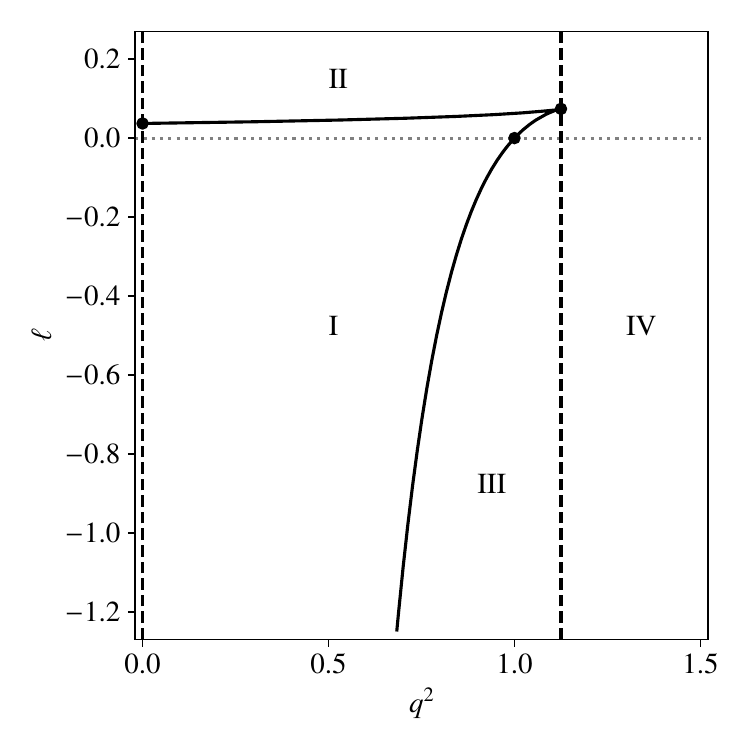}
	\caption{Phase diagram for ECPOs of the RN(a)dS spacetime in the $( q^{2}, \ell )$-plane, where $q = Q/M$ is the rescaled charge parameter and $\ell = \Lambda M^{2}/3$ is the rescaled cosmological constant. The dotted horizontal line is $\ell = 0$, corresponding to the asymptotically flat RN spacetime; this separates the asymptotically dS ($\ell > 0$) and asymptotically adS ($\ell < 0$) families. The solid contour has equation $\mathcal{B}( q^{2}, \ell ) = 16 \ell^{2} q^{6} + 8 \ell q^{4} - 36 \ell q^{2} + q^{2} + 27 \ell - 1 = 0$. Intriguingly, this curve separates black holes (region I) from naked singularities (regions II--IV), and is an ECPO phase boundary. The dashed vertical lines are the phase boundaries $q^{2} = 0$ and $q^{2} = 9/8$. In region I, there is one unstable ECPO. In region III, there are two ECPOs; the innermost is stable. There are no ECPOs in region II or IV.}
	\label{fig:charge_cosmo_phase_diagram}
\end{figure}

In Fig.~\ref{fig:charge_cosmo_phase_diagram}, the solid black contour in the $( q^{2}, \ell )$-plane has equation $\mathcal{B} ( q^{2}, \ell  ) = 0$; this separates RN(a)dS black holes (region I) from naked singularities (regions II--IV). The curve $\ell = 0$, which corresponds to the asymptotically flat case, is shown as a dotted horizontal line; this separates solutions that are asymptotically dS ($\ell > 0$) from those that are asymptotically adS ($\ell < 0$).

The equations governing geodesic motion on the RN(a)dS spacetime are separable. Consider an equatorial ($\theta = \pi/2$) ray with conserved non-zero energy $E = -p_{t}$. (Here, $p_{\mu}$ denotes the four-momentum of the photon.) The radial null geodesic equation can be written in the form $\dot{r}^{2} = E^{2} \mathcal{R}(u)$, where an overdot denotes differentiation with respect to the affine parameter, and
\begin{equation}
	\label{eqn:ecpo_quartic}
	\mathcal{R}(u) = 1 - b^{2} ( q^{2} u^{4} - 2 u^{3} + u^{2} - \ell ) .
\end{equation}
Here, $b = - p_{\phi}/ M p_{t}$ is the impact parameter of the ray.

ECPOs of the RN(a)dS spacetime satisfy $\dot{r} = 0 = \ddot{r}$, i.e., $\mathcal{R}(u) = 0 = \mathcal{R}^{\prime}(u)$. An orbit is stable if $\mathcal{R}^{\prime \prime}(u) < 0$. Thus, the problem of classifying the ECPOs of the RN(a)dS spacetime reduces to that of classifying the repeated roots of the quartic \eqref{eqn:ecpo_quartic}.

We seek values of the impact parameter $b$ that correspond to repeated roots of the polynomial $\mathcal{R}(u)$; these can be found by setting $\Delta_{u} (\mathcal{R}) = 0$ and solving for $b$, where $\Delta_{u}$ denotes the discriminant with respect to $u$. Phase boundaries in the $( q^{2}, \ell )$-plane can then be obtained by looking at the discriminant of $\Delta_{u} (\mathcal{R})$ with respect to $b$. Remarkably, this factorises as
\begin{equation}
	\label{eqn:ECPO_discriminant}
	\Delta_{b} \left( \frac{\Delta_{u} ( \mathcal{R} )}{ 16 b^{6} ( b^{2} \ell + 1 )} \right) = 256 q^{6} ( 8 q^{2} - 9 )^{6} \mathcal{B} ( q^{2}, \ell ) ,
\end{equation}
where, intriguingly, $\mathcal{B} ( q^{2}, \ell )$ is the quantity introduced in Eq.~\eqref{eqn:horizon_discriminant}, whose sign determines whether the RN(a)dS solution describes a black hole or a naked singularity.

Phase boundaries in the $( q^{2}, \ell )$-plane are given by setting the right-hand side of Eq.~\eqref{eqn:ECPO_discriminant} to zero, which occurs when $q^{2} = 0$, $q^{2} = 9/8$, or $\mathcal{B} ( q^{2}, \ell ) = 0$. These phase boundaries intersect at the points $( q^{2}, \ell ) = ( 0, 1/27 )$ and $( q^{2}, \ell ) = ( 9/8, 2/27 )$.

Figure \ref{fig:charge_cosmo_phase_diagram} shows the phase diagram for ECPOs of the RN(a)dS family of spacetimes in the $( q^{2}, \ell )$-plane. As described above, region I corresponds to black holes, and regions II--IV correspond to naked singularities. In region I, there is one unstable ECPO; there are no ECPOs in region II or IV; region III admits a pair of ECPOs, the innermost of which is stable.

A more comprehensive study of null geodesic motion in the RN(a)dS spacetime is presented by Stuchl\'{i}k and Hled\'{i}k in Ref.~\cite{StuchlikHledik2002}. ECPOs in the more general Kerr--Newman--(a)dS [KN(a)dS] spacetimes, which include the RN(a)dS geometry as a special case, are studied in more detail in Ref.~\cite{StuchlikHledik2000}. For a discussion of the extremal KNadS case, see Ref.~\cite{TangOngWang2017}.

\section{Stationary axisymmetric spacetimes and Einstein--Maxwell field equations \label{sec:geometry}}

General relativity with a cosmological constant and an electromagnetic field is described by the action
\begin{equation}
	\label{eqn:em_action}
	S = \frac{1}{16 \pi} \int ( R - 2 \Lambda - F_{\mu \nu} F^{\mu \nu} ) \sqrt{-g} \, \ed^{4} x.
\end{equation}
Here, $g$ is the determinant of the spacetime metric $g_{\mu \nu}$; $R$ is the Ricci curvature scalar, which is the trace of the Ricci tensor $R_{\mu \nu}$; $\Lambda$ is the cosmological constant; and $F_{\mu \nu}$ is the Faraday tensor that describes the Maxwell electromagnetic field. The Faraday tensor can be expressed in terms of an electromagnetic one-form potential $A_{\mu}$ as $F_{\mu \nu} = \nabla_{\mu} A_{\nu} - \nabla_{\nu} A_{\mu} = \partial_{\mu} A_{\nu} - \partial_{\nu} A_{\mu}$, where $\nabla_{\mu}$ is the covariant derivative and $\partial_{\mu}$ the standard partial derivative with respect to the spacetime coordinates $x^{\mu}$.

Extremising the action \eqref{eqn:em_action} with respect to $g_{\mu \nu}$ yields the Einstein field equations
\begin{equation}
	\label{eqn:einstein}
	G_{\mu \nu} + \Lambda g_{\mu \nu} = 8 \pi T_{\mu \nu} ,
\end{equation}
where $G_{\mu \nu} = R_{\mu \nu} - \frac{1}{2} R \, g_{\mu \nu}$ is the Einstein tensor, and
\begin{equation}
	\label{eqn:em_set}
	T_{\mu \nu} = \frac{1}{4 \pi} \left( F_{\mu \alpha} F_{\nu}^{\phantom{\nu} \alpha} - \frac{1}{4} g_{\mu \nu} F_{\alpha \beta} F^{\alpha \beta} \right)
\end{equation}
is the electromagnetic stress--energy tensor. Similarly, extremising the action \eqref{eqn:em_action} with respect to $A_{\mu}$ yields the source-free ($J^{\mu} = 0$) Maxwell equations
\begin{equation}
	\label{eqn:maxwell}
	\nabla_{\mu} F^{\mu \nu} = 0 .
\end{equation}
Together, Eqs.~\eqref{eqn:einstein} and \eqref{eqn:maxwell} are referred to as the Einstein--Maxwell equations.

We note that the cosmological term $\Lambda g_{\mu \nu}$ that appears on the left-hand side of Eq.~\eqref{eqn:einstein} can be taken to the right-hand side and absorbed into the definition of the stress--energy tensor. The cosmological constant can therefore be viewed as a stress--energy source, with effective stress--energy tensor $\hat{T}^{\Lambda}_{\mu \nu} = - \frac{\Lambda}{8 \pi} g_{\mu \nu}$.

Let us restrict our attention to four-dimensional stationary axisymmetric gravitational and electromagnetic fields. In such contexts, the line element can be generically expressed in Weyl--Lewis--Papapetrou coordinates $\{t, \rho, z, \phi \}$ as
\begin{equation}
	\begin{split}
		\label{eqn:sa_metric}
		\ed s^{2} &= g_{\mu \nu} \, \ed x^{\mu} \ed x^{\nu} \\
		&= - U \, ( \ed t - \omega \, \ed \phi )^{2} + \frac{1}{U} \left[ e^{2 \gamma} ( \ed \rho^{2} + \ed z^{2} ) + B^{2} \ed \phi^{2} \right] ,
	\end{split}
\end{equation}
and the electromagnetic four-potential as
\begin{equation}
	\label{eqn:sa_em_potential}
	A = A_{\mu} \, \ed x^{\mu} = A_{t} \, \ed t + A_{\phi} \, \ed \phi .
\end{equation}
Here, all of the metric components $g_{\mu \nu}$ and electromagnetic potentials $A_{\mu}$ are functions of $\rho$ and $z$ only. The spacetime \eqref{eqn:sa_metric} admits a pair of Killing vectors $\partial_{t}$ and $\partial_{\phi}$, which encode the stationary and axial symmetries, respectively.

In this work, we introduce the following nomenclature (all in the absence of electromagnetic sources, i.e., $J^{\mu} = 0$): \emph{pure vacuum} (or, simply, \emph{vacuum}) describes scenarios with $F_{\mu \nu} = 0$ and $\Lambda = 0$; \emph{electrovacuum} is used for $F_{\mu \nu} \neq 0$ and $\Lambda = 0$; \emph{$\Lambda$-vacuum} describes spacetimes with $F_{\mu \nu} = 0$ and  $\Lambda \neq 0$; and \emph{$\Lambda$-electrovacuum} spacetimes have $F_{\mu \nu} \neq 0$ and $\Lambda \neq 0$.

When $\Lambda \neq 0$, the metric component $g_{\phi \phi}$, which features the function $B$, is more general than the familiar form it takes in (electro)vacuum, where one can set $B = \rho$ without loss of generality. This is discussed in more detail below. In $\Lambda$-electrovacuum, we take $B$ to be non-negative by convention. Furthermore, we note that the line element \eqref{eqn:sa_metric} is appropriate for four-dimensional stationary axisymmetric solutions with a cosmological constant and an electromagnetic field; however, a more general form may be required when considering arbitrary matter sources \cite{Astorino2012}.

Astorino \cite{Astorino2012} presents a formulation of the Einstein--Maxwell equations \eqref{eqn:einstein} and \eqref{eqn:maxwell} for stationary axisymmetric potentials and spacetimes with a cosmological constant. For our purposes, the relevant Einstein field equations, expressed in terms of the functions $\{ U, \omega, \gamma, B, A_{t}, A_{\phi} \}$, are
\begin{align}
	0 &= U e^{-2 \gamma} \bnab^{2} B + 2 \Lambda B ,
	\label{eqn:EFE_B}
	\\
	0 &= \bnab \cdot \left( \frac{U^{2}}{B} \bnab \omega \right)
	+ \frac{4 U}{B} \left[ \omega \left( \bnab A_{t} \right)^{2} - 		\bnab A_{t} \cdot \bnab A_{\phi} \right] ,
	\label{eqn:EFE_omega} 
	\\
	\begin{split}
		0 &= \frac{U^{4}}{B} \left( \bnab \omega \right)^{2} + U \bnab \cdot \left( B \bnab U \right) - U^{2} \bnab^{2} B
		- B \left( \bnab U \right)^{2}  \\
		&\qquad - \frac{2 U^{3}}{B} \left[ \omega^{2} \left( \bnab A_{t} \right)^{2} + ( \bnab A_{\phi} )^{2} \right] \\
		&\qquad  + \frac{4 U^{3} \omega}{B}  \bnab A_{t} \cdot \bnab A_{\phi} - 2 B U \left( \bnab A_{t} \right)^{2} .
		\label{eqn:EFE_U}
	\end{split}
\end{align}

Here, $\bnab = ( \partial_{\rho}, \partial_{z} )$ is the standard two-gradient operator on the embedded space with metric $\ed \sigma^{2} = \ed \rho^{2} + \ed z^{2}$, where $\rho$ and $z$ are to be thought of as Cartesian coordinates. In particular, $\bnab \psi = ( \partial_{\rho} \psi, \partial_{z} \psi )$ and $\bnab^{2} \psi = \partial_{\rho}^{2} \psi + \partial_{z}^{2} \psi$ for an arbitrary scalar function $\psi = \psi (\rho, z )$, and $\bnab \cdot \mathbf{v} = \partial_{\rho} \varv_{\rho} + \partial_{z} \varv_{z}$ for an arbitrary vector field $\mathbf{v} = ( \varv_{\rho} (\rho, z ), \varv_{z} (\rho, z ) )$. We note that this differs from the standard notation used by Ernst \cite{Ernst1968a, Ernst1968b} and other authors (e.g.~\cite{DolanShipley2016}), who use a cylindrical gradient operator. In the Appendix, we describe how to map between the field equations presented here and those of Ernst \cite{Ernst1968b}.

In (electro)vacuum, where $\Lambda = 0$, the field equation \eqref{eqn:EFE_B} reduces to $\bnab^{2} B = 0$, so $B$ must be a harmonic function. One can always choose a system of harmonic coordinates $\left\{ \bar{\rho}, \bar{z} \right\}$ such that $B = \bar{\rho}$. In harmonic coordinates, the line element \eqref{eqn:sa_metric} reduces to the more familiar Weyl line element, and the field equations \eqref{eqn:EFE_omega} and \eqref{eqn:EFE_U} reduce to the (electro)vacuum Ernst equations \cite{Ernst1968a,Ernst1968b}.

However, when $\Lambda \neq 0$, $B$ satisfies the Helmholtz-type equation \eqref{eqn:EFE_B}, and it is not possible to set $B = \rho$ without loss of generality. This demonstrates the importance of keeping the function $B$ in the line element \eqref{eqn:sa_metric} in situations with a non-zero cosmological constant. The $\Lambda$-vacuum case is discussed in more detail by Suvorov and Melatos \cite{SuvorovMelatos2016} as part of a more general treatment of stationary axisymmetric fields in $f(R)$ gravity. The field equations presented here with $A_{\mu} = 0$ are equivalent to those of Ref.~\cite{SuvorovMelatos2016} with $f(R) = R - 2 \Lambda$, i.e., general relativity with a cosmological constant but no stress--energy sources; see the Appendix for details.

\section{Light rings and their stability \label{sec:light_rings}}

\subsection{Hamiltonian formalism for null geodesics \label{sec:hamiltonian}}

The geodesics $x^{\mu}(\lambda)$ of the spacetime \eqref{eqn:sa_metric} are the integral curves of Hamilton's equations with Hamiltonian function $H = \frac{1}{2} g^{\mu \nu} p_{\mu} p_{\nu}$, where $p_{\mu} = g_{\mu \nu} \dot{x}^{\nu}$ are the canonical momenta, and an overdot denotes differentiation with respect to the affine parameter $\lambda$.

The Hamiltonian $H$ and the momenta $p_{t}$ and $p_{\phi}$ are conserved along geodesics, with $H = 0$ for null rays. (These constants of motion correspond, respectively, to the trivial rank-two Killing tensor $g_{\mu \nu}$, and the Killing vectors $\partial_{t}$ and $\partial_{\phi}$ of the spacetime.) Moreover, in the null case, one may set $p_{t} = -1$ without loss of generality by availing the affine-rescaling freedom ($\lambda \mapsto a \lambda$, $a \in \mathbb{R}$).

Null geodesics are invariant under conformal transformations of the metric tensor. Applying the transformation $g_{\mu \nu} \mapsto U e^{-2 \gamma} g_{\mu \nu}$, we may recast the Hamiltonian in canonical form as
\begin{align}
	H &= \frac{1}{2} ( p_{\rho}^{2} + p_{z}^{2} ) + V , \label{eqn:hamiltonian} \\
	V &= - \frac{e^{2 \gamma}}{2 B^{2}} \left[ \left( \frac{B^{2}}{U^{2}} - \omega^{2} \right) + 2 \omega p_{\phi} - p_{\phi}^{2} \right] . 
	\label{eqn:potential}
\end{align}
Following the approach laid out in Ref.~\cite{DolanShipley2016}, we introduce a pair of effective potentials
\begin{equation}
	\label{eqn:effective_potentials}
	h^{\pm} ( \rho, z ) = \frac{B}{U} \pm \omega ,
\end{equation}
which are independent of the parameter $p_{\phi}$. This permits us to factorise the geodesic potential \eqref{eqn:potential} as
\begin{equation}
	\label{eqn:potential_fact}
	V = - \frac{e^{2 \gamma}}{2 B^{2}} ( h^{+} - p_{\phi} ) ( h^{-} + p_{\phi} ) .
\end{equation}
In the static case, $\omega = 0$ and we write $h = h^{\pm}$. In electrovacuum ($\Lambda = 0$), we have $B = \rho$, and the Hamiltonian \eqref{eqn:hamiltonian} and effective potentials \eqref{eqn:effective_potentials} reduce to those presented in Ref.~\cite{DolanShipley2016}.

\subsection{Classification of light rings \label{sec:classification}}

Light rings are circular null geodesics of constant $\rho$ and $z$, i.e., they satisfy $V = 0$ and $\bnab V = \mathbf{0}$. These existence conditions for light rings can be expressed in terms of the effective potentials \eqref{eqn:effective_potentials} as $h^{\pm} = \pm p_{\phi}$ and $\bnab h^{\pm} = \mathbf{0}$. In other words, light rings are fixed points of $h^{\pm}$.

The stability of light rings can be determined by considering the trace and determinant of $\mathcal{H}(h^{\pm})$, the Hessian matrix of second-order partial derivatives. Stable light rings correspond to minima of the geodesic potential \eqref{eqn:potential_fact}, and therefore maxima of $h^{\pm}$: a light ring is stable if $\det{\mathcal{H}(h^{\pm})} > 0$ and $\tr{\mathcal{H}(h^{\pm})} < 0$.

Here, we demonstrate how the field equations \eqref{eqn:EFE_B}--\eqref{eqn:EFE_U} may be used to classify light rings in stationary axisymmetric spacetimes with an electromagnetic field and a cosmological constant.

The first and second partial derivatives of the potentials \eqref{eqn:effective_potentials} with respect to $\rho$ are given by
\begin{align}
	\partial_{\rho} h^{\pm} &= \frac{1}{U} \partial_{\rho} B - \frac{B}{U^{2}} \partial_{\rho} U \pm \partial_{\rho} \omega , \\
	\begin{split}
		\partial_{\rho}^{2} h^{\pm} &= \frac{1}{U} \partial_{\rho}^{2} B - \frac{2}{U^{2}} ( \partial_{\rho} B ) ( \partial_{\rho} U ) + \frac{2 B}{U^{3}} ( \partial_{\rho} U )^{2} \\
		& \qquad - \frac{B}{U^{2}} \partial_{\rho}^{2} U \pm \partial_{\rho}^{2} \omega .
	\end{split}
\end{align}
The partial derivatives with respect to $z$ can be found straightforwardly by symmetry. Using these expressions, we find
\begin{equation}
	\label{eqn:trace_hessian}
	\begin{split}
		\tr{\mathcal{H}(h^{\pm})} &= \partial_{\rho}^{2} h^{\pm} + \partial_{z}^{2} h^{\pm} \\
		&= \frac{1}{U} \bnab^{2} B - \frac{2}{U^{2}} \bnab B \cdot \bnab U + \frac{2 B}{U^{3}} \left( \bnab U \right)^{2} \\
		&\qquad - \frac{B}{U^{2}} \bnab^{2} U \pm \bnab^{2} \omega .
	\end{split}
\end{equation}

Expanding and rearranging Eq.~\eqref{eqn:EFE_U}, we can express the Laplacian of $U$ as
\begin{equation}
	\label{eqn:laplacian_U}
	\begin{split}
	\bnab^{2} U &= \frac{U}{B} \bnab^{2} B + \frac{1}{U} \left( \bnab U \right)^{2} - \frac{1}{B} \bnab B \cdot \bnab U - \frac{U^{3}}{B} \left( \bnab \omega \right)^{2} \\
	& \qquad  + \frac{2 U^{2}}{B^{2}} \left[ \omega^{2} \left( \bnab A_{t} \right)^{2} + ( \bnab A_{\phi} )^{2} \right] \\
	&\qquad  - \frac{4 U^{2} \omega}{B^{2}}  \bnab A_{t} \cdot \bnab A_{\phi} + 2 \left( \bnab A_{t} \right)^{2} .
\end{split}
\end{equation}
Doing the same with Eq.~\eqref{eqn:EFE_omega} gives
\begin{equation}
	\label{eqn:laplacian_omega}
	\begin{split}
		\bnab^{2} \omega &= \frac{1}{B} \bnab B \cdot \bnab \omega - \frac{2}{U} \bnab U \cdot \bnab \omega \\
		&\qquad - \frac{4}{U} \left[ \omega \left( \bnab A_{t} \right)^{2} - \bnab A_{t} \cdot \bnab A_{\phi} \right] .
	\end{split}
\end{equation}

Now, we may use Eqs.~\eqref{eqn:laplacian_U} and \eqref{eqn:laplacian_omega} to replace terms involving $\bnab^{2} U$ and $\bnab^{2} \omega$ in Eq.~\eqref{eqn:trace_hessian}. Upon simplification, we find
\begin{equation}
	\label{eqn:trace_hessian_2}
	\begin{split}
	\tr{\mathcal{H}(h^{\pm})} &= \frac{B}{U^{3}} \left( \bnab U \right)^{2} - \frac{1}{U^{2}} \bnab B \cdot \bnab U + \frac{U}{B} \left( \bnab \omega \right)^{2} \\
	&\qquad \pm \left( \frac{1}{B} \bnab B \cdot \bnab \omega - \frac{2}{U} \bnab U \cdot \bnab \omega \right) \\
	&\qquad - \frac{2}{B} ( h^{\pm} \bnab A_{t} \mp \bnab A_{\phi} )^{2}	.
\end{split}
\end{equation}
Intriguingly, terms involving $\bnab^{2} B$ vanish from the right-hand side of Eq.~\eqref{eqn:trace_hessian_2}, so there is no need to avail the remaining field equation \eqref{eqn:EFE_B}. As a result, we do not introduce the cosmological constant $\Lambda$ (or the function $\gamma$ that appears in the metric) into the expression for the trace of the Hessian. Moreover, Eq.~\eqref{eqn:trace_hessian_2} now involves only first derivatives of the metric functions $U$, $\omega$ and $B$, and the electromagnetic potentials $A_{t}$ and $A_{\phi}$.

As outlined above, light rings are stationary points of $h^{\pm}$, satisfying $\bnab h^{\pm} = \mathbf{0}$. Rearranging this equality, we see that
\begin{equation}
	\label{eqn:grad_omega}
	\pm \bnab \omega = \frac{B}{U^{2}} \bnab U - \frac{1}{U} \bnab B 
\end{equation}
must hold at fixed points of $h^{\pm}$. Squaring both sides of Eq.~\eqref{eqn:grad_omega}, we see that the following must also hold:
\begin{equation}
	\label{eqn:grad_omega_squared}
	\left( \bnab \omega \right)^{2} = \frac{B^{2}}{U^{4}} \left( \bnab U \right)^{2} - \frac{2 B}{U^{3}} \bnab B \cdot \bnab U + \frac{1}{U^{2}} \left( \bnab B \right)^{2} .
\end{equation}

Using Eqs.~\eqref{eqn:grad_omega} and \eqref{eqn:grad_omega_squared} to replace terms involving $\bnab \omega$ on the right-hand side of Eq.~\eqref{eqn:trace_hessian_2}, a remarkable cancellation occurs and we are left with
\begin{equation}
	\label{eqn:trace_hessian_3}
	\tr{\mathcal{H}(h^{\pm})} = - \frac{2}{B} ( h^{\pm} \bnab A_{t} \mp \bnab A_{\phi} )^{2} .
\end{equation}

When $\Lambda = 0$, i.e., in (electro)vacuum, one can set $B = \rho$, and Eq.~\eqref{eqn:trace_hessian_3} reduces to the key result of Ref.~\cite{DolanShipley2016}: in the absence of a cosmological constant, stable photon orbits cannot be ruled out in electrovacuum scenarios where at least one of $\bnab A_{t}$ and $\bnab A_{\phi}$ is non-zero; however, they are forbidden in pure vacuum.

In $\Lambda$-electrovacuum, if at least one of $\bnab A_{t}$ and $\bnab A_{\phi}$ is non-zero, then $\tr{\mathcal{H}(h^{\pm})} < 0$ by Eq.~\eqref{eqn:trace_hessian_3}. Therefore, local minima of $h^{\pm}$ are forbidden, but local maxima, which exist if $\det{\mathcal{H}(h^{\pm})} > 0$, cannot be ruled out. Crucially, this shows that a non-zero cosmological constant cannot be introduced to forbid the existence of stable light rings in stationary axisymmetric spacetimes with an electromagnetic field.

However, in $\Lambda$-vacuum, one can clearly see from the right-hand side of Eq.~\eqref{eqn:trace_hessian_3} that $\tr{\mathcal{H}(h^{\pm})}$ vanishes at a light ring, so $\det{\mathcal{H}(h^{\pm})} = - \left[ ( \partial_{\rho}^{2} h^{\pm} )^{2} + ( \partial_{\rho} \partial_{z} h^{\pm} )^{2} \right] \leq 0$. It follows that extrema of $h^{\pm}$ and therefore stable light rings are forbidden, as is the case in pure vacuum.

\section{Discussion \label{sec:discussion}}

In this work, we have used a Hamiltonian formalism (Sec.~\ref{sec:hamiltonian}) to describe light rings in four spacetime dimensions under the general assumptions of stationarity and axisymmetry. Using a subset of the $\Lambda$-electrovacuum field equations (Sec.~\ref{sec:geometry}), we have classified the stability of light rings by considering the fixed points of a pair of two-dimensional effective potentials (Sec.~\ref{sec:classification}).

Our results show that generic stable light rings are forbidden in $\Lambda$-vacuum, but cannot be ruled out in $\Lambda$-electrovacuum thanks to the form of Eq.~\eqref{eqn:trace_hessian_3}. This extends the main conclusion of Ref.~\cite{DolanShipley2016}, in which it was demonstrated that stable light rings are not permitted in pure vacuum but may arise in electrovacuum.

Hence, we have demonstrated that the cosmological constant cannot be introduced to the Einstein--Maxwell equations to rescue us from the existence of stable light rings, and therefore spacetime instabilities, in four-dimensional stationary axisymmetric contexts. Viewed from a different angle, our results show that the cosmological constant alone is not a mechanism for the existence of stable light rings, but the electromagnetic field can, in principle, give rise to such orbits.

A number of analyses have uncovered stable light rings in a variety of other settings (e.g. around horizonless ultra-compact objects \cite{CardosoEtAl2014, CunhaEtAl2016, CunhaBertiHerdeiro2017, CunhaHerdeiro2020}), which suggests that an electromagnetic field is not a necessary requirement for the existence of stable light rings. To generalise this work, it is natural to ask which matter sources besides the electromagnetic field can, in principle, give rise to stable light rings. This would involve using the method presented in Secs.~\ref{sec:geometry} and \ref{sec:light_rings} with more general stress--energy terms in the action \eqref{eqn:em_action} and field equations \eqref{eqn:einstein}. We note that a more general line element than \eqref{eqn:sa_metric} may be required when considering other matter fields. Changing the form of the line element would consequently alter the Hamiltonian formalism presented in Sec.~\ref{sec:hamiltonian}.

Furthermore, one could use our approach to study the existence and stability of light rings in modified theories of gravity. As a starting point, one might consider $f(R)$ theories, in which the term $R - 2 \Lambda$ in the integrand of the action \eqref{eqn:em_action} is replaced by a general function of the Ricci scalar $f(R)$. An Ernst-like formulation of the field equations of $f(R)$ gravity is presented by Suvorov and Melatos \cite{SuvorovMelatos2016} in the absence of electromagnetic fields ($F_{\mu \nu} = 0$). Even in this relatively simple case, the field equations are much less tractable than Eqs.~\eqref{eqn:EFE_B}--\eqref{eqn:EFE_U} of this work.

Recent analyses have been dedicated to the study of \emph{fundamental photon orbits} \cite{CunhaHerdeiroRadu2017, ShipleyDolan2016, CunhaHerdeiro2018, LimaJuniorEtAl2021}, which are generalisations of the spherical photon orbits of Kerr spacetime \cite{Teo2003} to stationary axisymmetric spacetimes whose geodesic motion is not necessarily integrable. Such orbits -- both stable and unstable -- are studied in the context of Kerr--Newman--NUT--(a)dS black holes in Ref.~\cite{GrenzebachPerlickLammerzahl2014}.

We caution that these more general fundamental photon orbits fall beyond the classification presented in Sec.~\ref{sec:light_rings}, which applies only to light rings. An alternative approach would therefore be required to provide a more complete description of null geodesic motion -- covering both light rings and fundamental photon orbits -- in four-dimensional stationary axisymmetric spacetimes.

\begin{acknowledgments}
	With thanks to Sam Dolan for helpful discussions and feedback.
\end{acknowledgments}

\appendix*

\section{\uppercase{Einstein--Maxwell equations} \label{sec:em_equations}}

In Ref.~\cite{Astorino2012}, Astorino expresses the metric for a four-dimensional stationary axisymmetric $\Lambda$-electrovacuum spacetime in the form $\ed s^{2} = - \alpha e^{\Omega/2} ( \ed t - \omega \, \ed \phi )^{2} + \alpha e^{-\Omega/2} \, \ed \phi^{2} + \alpha^{-1/2} e^{2 \nu} ( \ed \rho^{2} + \ed z^{2} )$; see also Ref.~\cite{CharmousisEtAl2007}. (We note that there is a sign error in the coefficient of $\ed t \, \ed \phi$ in Eq.~(2.2) of Ref.~\cite{Astorino2012}.) This can be mapped to our line element \eqref{eqn:sa_metric} via the following transformation:
\begin{equation}
	\left\{ e^{\Omega/2} , \omega, e^{2 \nu} , \alpha \right\} \mapsto \left\{ \frac{U}{B}, \omega, \frac{e^{2 \gamma} \sqrt{B}}{U}, B \right\} .
\end{equation}

Applying this transformation to Eqs.~(2.5)--(2.10) of Ref.~\cite{Astorino2012} yields the Einstein--Maxwell equations for the spacetime \eqref{eqn:sa_metric} with electromagnetic potential \eqref{eqn:sa_em_potential}. In particular, Eqs.~\eqref{eqn:EFE_B} and \eqref{eqn:EFE_omega} of this work are equivalent to Eqs.~(2.5) and (2.6), respectively, of Ref.~\cite{Astorino2012}. The third field equation that we make use of -- Eq.~\eqref{eqn:EFE_U} -- can be obtained by inserting Eq.~(2.6) into Eq.~(2.7) of Ref.~\cite{Astorino2012}, and using the identity $\bnab \Omega = e^{-\Omega} \bnab e^{\Omega}$. The remaining Einstein--Maxwell equations presented in Ref.~\cite{Astorino2012} are not required in this work.

In the absence of a cosmological constant, the field equations \eqref{eqn:EFE_B}--\eqref{eqn:EFE_U} reduce to the electrovacuum equations of Ernst \cite{Ernst1968b}, used in Ref.~\cite{DolanShipley2016}. This can be shown by setting $\Lambda = 0$ and $B = \rho$, and by transforming the gradient operator $\bnab$ used in this work to the one used by Ernst, which we denote by $\tilde{\bnab}$. These derivative operators are related via $\bnab \psi = \tilde{\bnab} \psi$, $\bnab^{2} \psi = \tilde{\bnab}^{2} \psi - \frac{1}{\rho} \partial_{\rho} \psi$, $\bnab \cdot \mathbf{v} = \tilde{\bnab} \cdot \mathbf{v} - \frac{1}{\rho} \varv_{\rho}$, where $\psi ( \rho, z )$ is an arbitrary scalar field, and $\mathbf{v} = ( \varv_{\rho} ( \rho, z ), \varv_{z} ( \rho, z ) )$ an arbitrary vector field. (We note also that the function $U$ that appears throughout this work is denoted by $f$ in Ernst's work \cite{Ernst1968a,Ernst1968b}.)

In $\Lambda$-vacuum, we may contrast the field equations \eqref{eqn:EFE_B}--\eqref{eqn:EFE_U} with those of Suvorov and Melatos \cite{SuvorovMelatos2016}, who present an Ernst formulation of the field equations in $f(R)$ gravity with no matter fields. General relativity with a cosmological constant is recovered when $f(R) = R - 2 \Lambda$. With $A_{t} = 0 = A_{\phi}$, Eqs.~\eqref{eqn:EFE_B}--\eqref{eqn:EFE_U} are equivalent to Eq.~(13) and the real and imaginary parts of Eq.~(18) with $f(R) = R - 2 \Lambda$ in Ref.~\cite{SuvorovMelatos2016}. As outlined in Sec.~\ref{sec:discussion}, extending this analysis to $f(R)$ theories (or other modified theories) of gravity would be a natural step towards providing a more complete description of the existence and stability of light rings in stationary axisymmetric gravitational fields.

\bibliographystyle{apsrev4-2}
\bibliography{Shipley_StablePhotonOrbitsLambdaElectrovacuum_refs}

\begin{thebibliography}{31}%
\makeatletter
\providecommand \@ifxundefined [1]{%
 \@ifx{#1\undefined}
}%
\providecommand \@ifnum [1]{%
 \ifnum #1\expandafter \@firstoftwo
 \else \expandafter \@secondoftwo
 \fi
}%
\providecommand \@ifx [1]{%
 \ifx #1\expandafter \@firstoftwo
 \else \expandafter \@secondoftwo
 \fi
}%
\providecommand \natexlab [1]{#1}%
\providecommand \enquote  [1]{``#1''}%
\providecommand \bibnamefont  [1]{#1}%
\providecommand \bibfnamefont [1]{#1}%
\providecommand \citenamefont [1]{#1}%
\providecommand \href@noop [0]{\@secondoftwo}%
\providecommand \href [0]{\begingroup \@sanitize@url \@href}%
\providecommand \@href[1]{\@@startlink{#1}\@@href}%
\providecommand \@@href[1]{\endgroup#1\@@endlink}%
\providecommand \@sanitize@url [0]{\catcode `\\12\catcode `\$12\catcode
  `\&12\catcode `\#12\catcode `\^12\catcode `\_12\catcode `\%12\relax}%
\providecommand \@@startlink[1]{}%
\providecommand \@@endlink[0]{}%
\providecommand \url  [0]{\begingroup\@sanitize@url \@url }%
\providecommand \@url [1]{\endgroup\@href {#1}{\urlprefix }}%
\providecommand \urlprefix  [0]{URL }%
\providecommand \Eprint [0]{\href }%
\providecommand \doibase [0]{https://doi.org/}%
\providecommand \selectlanguage [0]{\@gobble}%
\providecommand \bibinfo  [0]{\@secondoftwo}%
\providecommand \bibfield  [0]{\@secondoftwo}%
\providecommand \translation [1]{[#1]}%
\providecommand \BibitemOpen [0]{}%
\providecommand \bibitemStop [0]{}%
\providecommand \bibitemNoStop [0]{.\EOS\space}%
\providecommand \EOS [0]{\spacefactor3000\relax}%
\providecommand \BibitemShut  [1]{\csname bibitem#1\endcsname}%
\let\auto@bib@innerbib\@empty
\bibitem [{\citenamefont {Teo}(2003)}]{Teo2003}%
  \BibitemOpen
  \bibfield  {author} {\bibinfo {author} {\bibfnamefont {E.}~\bibnamefont
  {Teo}},\ }\href {https://doi.org/10.1023/A:1026286607562} {\bibfield
  {journal} {\bibinfo  {journal} {{Gen. Relativ. Gravit.}}\ }\textbf {\bibinfo
  {volume} {35}},\ \bibinfo {pages} {1909} (\bibinfo {year}
  {2003})}\BibitemShut {NoStop}%
\bibitem [{\citenamefont {Perlick}(2004)}]{Perlick2004}%
  \BibitemOpen
  \bibfield  {author} {\bibinfo {author} {\bibfnamefont {V.}~\bibnamefont
  {Perlick}},\ }\href {https://doi.org/10.12942/lrr-2004-9} {\bibfield
  {journal} {\bibinfo  {journal} {{Living Rev. Relativity}}\ }\textbf {\bibinfo
  {volume} {7}},\ \bibinfo {pages} {1} (\bibinfo {year} {2004})},\ \Eprint
  {https://arxiv.org/abs/1010.3416} {arXiv:1010.3416 [gr-qc]} \BibitemShut
  {NoStop}%
\bibitem [{\citenamefont {Cunha}\ and\ \citenamefont
  {Herdeiro}(2018)}]{CunhaHerdeiro2018}%
  \BibitemOpen
  \bibfield  {author} {\bibinfo {author} {\bibfnamefont {P.~V.~P.}\
  \bibnamefont {Cunha}}\ and\ \bibinfo {author} {\bibfnamefont {C.~A.~R.}\
  \bibnamefont {Herdeiro}},\ }\href {https://doi.org/10.1007/s10714-018-2361-9}
  {\bibfield  {journal} {\bibinfo  {journal} {{Gen. Relativ. Gravit.}}\
  }\textbf {\bibinfo {volume} {50}},\ \bibinfo {pages} {1} (\bibinfo {year}
  {2018})},\ \Eprint {https://arxiv.org/abs/1801.00860} {arXiv:1801.00860
  [gr-qc]} \BibitemShut {NoStop}%
\bibitem [{\citenamefont {Liang}(1974)}]{Liang1974}%
  \BibitemOpen
  \bibfield  {author} {\bibinfo {author} {\bibfnamefont {E.~P.~T.}\
  \bibnamefont {Liang}},\ }\href {https://doi.org/10.1103/PhysRevD.9.3257}
  {\bibfield  {journal} {\bibinfo  {journal} {{Phys. Rev. D}}\ }\textbf
  {\bibinfo {volume} {9}},\ \bibinfo {pages} {3257} (\bibinfo {year}
  {1974})}\BibitemShut {NoStop}%
\bibitem [{\citenamefont {Calvani}\ \emph {et~al.}(1980)\citenamefont
  {Calvani}, \citenamefont {de~Felice},\ and\ \citenamefont
  {Nobili}}]{CalvaniDeFeliceNobili1980}%
  \BibitemOpen
  \bibfield  {author} {\bibinfo {author} {\bibfnamefont {M.}~\bibnamefont
  {Calvani}}, \bibinfo {author} {\bibfnamefont {F.}~\bibnamefont {de~Felice}},\
  and\ \bibinfo {author} {\bibfnamefont {L.}~\bibnamefont {Nobili}},\ }\href
  {https://doi.org/10.1088/0305-4470/13/10/018} {\bibfield  {journal} {\bibinfo
   {journal} {{J. Phys. A}}\ }\textbf {\bibinfo {volume} {13}},\ \bibinfo
  {pages} {3213} (\bibinfo {year} {1980})}\BibitemShut {NoStop}%
\bibitem [{\citenamefont {Stuchl\'{i}k}(1981)}]{Stuchlik1981}%
  \BibitemOpen
  \bibfield  {author} {\bibinfo {author} {\bibfnamefont {Z.}~\bibnamefont
  {Stuchl\'{i}k}},\ }\href@noop {} {\bibfield  {journal} {\bibinfo  {journal}
  {Bull. Astron. Inst. Czech.}\ }\textbf {\bibinfo {volume} {32}},\ \bibinfo
  {pages} {366} (\bibinfo {year} {1981})}\BibitemShut {NoStop}%
\bibitem [{\citenamefont {Balek}\ \emph {et~al.}(1989)\citenamefont {Balek},
  \citenamefont {Bi\v{c}\'{a}k},\ and\ \citenamefont
  {Stuchl\'{i}k}}]{BalekBicakStuchlik1989}%
  \BibitemOpen
  \bibfield  {author} {\bibinfo {author} {\bibfnamefont {V.}~\bibnamefont
  {Balek}}, \bibinfo {author} {\bibfnamefont {J.}~\bibnamefont
  {Bi\v{c}\'{a}k}},\ and\ \bibinfo {author} {\bibfnamefont {Z.}~\bibnamefont
  {Stuchl\'{i}k}},\ }\href@noop {} {\bibfield  {journal} {\bibinfo  {journal}
  {Bull. Astron. Inst. Czech.}\ }\textbf {\bibinfo {volume} {40}},\ \bibinfo
  {pages} {133} (\bibinfo {year} {1989})}\BibitemShut {NoStop}%
\bibitem [{\citenamefont {Khoo}\ and\ \citenamefont {Ong}(2016)}]{KhooOng2016}%
  \BibitemOpen
  \bibfield  {author} {\bibinfo {author} {\bibfnamefont {F.~S.}\ \bibnamefont
  {Khoo}}\ and\ \bibinfo {author} {\bibfnamefont {Y.~C.}\ \bibnamefont {Ong}},\
  }\href {https://doi.org/10.1088/0264-9381/33/23/235002} {\bibfield  {journal}
  {\bibinfo  {journal} {Classical Quantum Gravity}\ }\textbf {\bibinfo {volume}
  {33}},\ \bibinfo {pages} {235002} (\bibinfo {year} {2016})},\ \Eprint
  {https://arxiv.org/abs/1605.05774} {arXiv:1605.05774 [gr-qc]} \BibitemShut
  {NoStop}%
\bibitem [{\citenamefont {Dolan}\ and\ \citenamefont
  {Shipley}(2016)}]{DolanShipley2016}%
  \BibitemOpen
  \bibfield  {author} {\bibinfo {author} {\bibfnamefont {S.~R.}\ \bibnamefont
  {Dolan}}\ and\ \bibinfo {author} {\bibfnamefont {J.~O.}\ \bibnamefont
  {Shipley}},\ }\href {https://doi.org/10.1103/PhysRevD.94.044038} {\bibfield
  {journal} {\bibinfo  {journal} {{Phys. Rev. D}}\ }\textbf {\bibinfo {volume}
  {94}},\ \bibinfo {pages} {044038} (\bibinfo {year} {2016})},\ \Eprint
  {https://arxiv.org/abs/1605.07193} {arXiv:1605.07193 [gr-qc]} \BibitemShut
  {NoStop}%
\bibitem [{\citenamefont {Shipley}(2019)}]{Shipley2019}%
  \BibitemOpen
  \bibfield  {author} {\bibinfo {author} {\bibfnamefont {J.~O.}\ \bibnamefont
  {Shipley}},\ }\emph {\bibinfo {title} {Strong-field gravitational lensing by
  black holes}},\ \href {https://etheses.whiterose.ac.uk/24823/} {Ph.D.
  thesis},\ \bibinfo  {school} {University of Sheffield} (\bibinfo {year}
  {2019}),\ \Eprint {https://arxiv.org/abs/1909.04691} {arXiv:1909.04691
  [gr-qc]} \BibitemShut {NoStop}%
\bibitem [{\citenamefont {Cardoso}\ \emph {et~al.}(2014)\citenamefont
  {Cardoso}, \citenamefont {Crispino}, \citenamefont {Macedo}, \citenamefont
  {Okawa},\ and\ \citenamefont {Pani}}]{CardosoEtAl2014}%
  \BibitemOpen
  \bibfield  {author} {\bibinfo {author} {\bibfnamefont {V.}~\bibnamefont
  {Cardoso}}, \bibinfo {author} {\bibfnamefont {L.~C.~B.}\ \bibnamefont
  {Crispino}}, \bibinfo {author} {\bibfnamefont {C.~F.~B.}\ \bibnamefont
  {Macedo}}, \bibinfo {author} {\bibfnamefont {H.}~\bibnamefont {Okawa}},\ and\
  \bibinfo {author} {\bibfnamefont {P.}~\bibnamefont {Pani}},\ }\href
  {https://doi.org/10.1103/PhysRevD.90.044069} {\bibfield  {journal} {\bibinfo
  {journal} {{Phys. Rev. D}}\ }\textbf {\bibinfo {volume} {90}},\ \bibinfo
  {pages} {044069} (\bibinfo {year} {2014})},\ \Eprint
  {https://arxiv.org/abs/1406.5510} {arXiv:1406.5510 [gr-qc]} \BibitemShut
  {NoStop}%
\bibitem [{\citenamefont {Cunha}\ \emph {et~al.}(2016)\citenamefont {Cunha},
  \citenamefont {Grover}, \citenamefont {Herdeiro}, \citenamefont {Radu},
  \citenamefont {R\'{u}narsson},\ and\ \citenamefont {Wittig}}]{CunhaEtAl2016}%
  \BibitemOpen
  \bibfield  {author} {\bibinfo {author} {\bibfnamefont {P.~V.~P.}\
  \bibnamefont {Cunha}}, \bibinfo {author} {\bibfnamefont {J.}~\bibnamefont
  {Grover}}, \bibinfo {author} {\bibfnamefont {C.}~\bibnamefont {Herdeiro}},
  \bibinfo {author} {\bibfnamefont {E.}~\bibnamefont {Radu}}, \bibinfo {author}
  {\bibfnamefont {H.}~\bibnamefont {R\'{u}narsson}},\ and\ \bibinfo {author}
  {\bibfnamefont {A.}~\bibnamefont {Wittig}},\ }\href
  {https://doi.org/10.1103/PhysRevD.94.104023} {\bibfield  {journal} {\bibinfo
  {journal} {{Phys. Rev. D}}\ }\textbf {\bibinfo {volume} {94}},\ \bibinfo
  {pages} {104023} (\bibinfo {year} {2016})},\ \Eprint
  {https://arxiv.org/abs/1609.01340} {arXiv:1609.01340 [gr-qc]} \BibitemShut
  {NoStop}%
\bibitem [{\citenamefont {Cunha}\ \emph
  {et~al.}(2017{\natexlab{a}})\citenamefont {Cunha}, \citenamefont {Berti},\
  and\ \citenamefont {Herdeiro}}]{CunhaBertiHerdeiro2017}%
  \BibitemOpen
  \bibfield  {author} {\bibinfo {author} {\bibfnamefont {P.~V.~P.}\
  \bibnamefont {Cunha}}, \bibinfo {author} {\bibfnamefont {E.}~\bibnamefont
  {Berti}},\ and\ \bibinfo {author} {\bibfnamefont {C.~A.~R.}\ \bibnamefont
  {Herdeiro}},\ }\href {https://doi.org/10.1103/PhysRevLett.119.251102}
  {\bibfield  {journal} {\bibinfo  {journal} {{Phys. Rev. Lett.}}\ }\textbf
  {\bibinfo {volume} {119}},\ \bibinfo {pages} {251102} (\bibinfo {year}
  {2017}{\natexlab{a}})},\ \Eprint {https://arxiv.org/abs/1708.04211}
  {arXiv:1708.04211 [gr-qc]} \BibitemShut {NoStop}%
\bibitem [{\citenamefont {Cunha}\ and\ \citenamefont
  {Herdeiro}(2020)}]{CunhaHerdeiro2020}%
  \BibitemOpen
  \bibfield  {author} {\bibinfo {author} {\bibfnamefont {P.~V.~P.}\
  \bibnamefont {Cunha}}\ and\ \bibinfo {author} {\bibfnamefont {C.~A.~R.}\
  \bibnamefont {Herdeiro}},\ }\href
  {https://doi.org/10.1103/PhysRevLett.124.181101} {\bibfield  {journal}
  {\bibinfo  {journal} {{Phys. Rev. Lett.}}\ }\textbf {\bibinfo {volume}
  {124}},\ \bibinfo {pages} {181101} (\bibinfo {year} {2020})},\ \Eprint
  {https://arxiv.org/abs/2003.06445} {arXiv:2003.06445 [gr-qc]} \BibitemShut
  {NoStop}%
\bibitem [{\citenamefont {Ghosh}\ and\ \citenamefont
  {Sarkar}(2021)}]{GhoshSarkar2021}%
  \BibitemOpen
  \bibfield  {author} {\bibinfo {author} {\bibfnamefont {R.}~\bibnamefont
  {Ghosh}}\ and\ \bibinfo {author} {\bibfnamefont {S.}~\bibnamefont {Sarkar}},\
  }\href {https://doi.org/10.1103/PhysRevD.104.044019} {\bibfield  {journal}
  {\bibinfo  {journal} {{Phys. Rev. D}}\ }\textbf {\bibinfo {volume} {104}},\
  \bibinfo {pages} {044019} (\bibinfo {year} {2021})},\ \Eprint
  {https://arxiv.org/abs/2107.07370} {arXiv:2107.07370 [gr-qc]} \BibitemShut
  {NoStop}%
\bibitem [{\citenamefont {Cunha}\ \emph {et~al.}(2023)\citenamefont {Cunha},
  \citenamefont {Herdeiro}, \citenamefont {Radu},\ and\ \citenamefont
  {Sanchis-Gual}}]{CunhaEtAl2023}%
  \BibitemOpen
  \bibfield  {author} {\bibinfo {author} {\bibfnamefont {P.~V.~P.}\
  \bibnamefont {Cunha}}, \bibinfo {author} {\bibfnamefont {C.}~\bibnamefont
  {Herdeiro}}, \bibinfo {author} {\bibfnamefont {E.}~\bibnamefont {Radu}},\
  and\ \bibinfo {author} {\bibfnamefont {N.}~\bibnamefont {Sanchis-Gual}},\
  }\href {https://doi.org/10.1103/PhysRevLett.130.061401} {\bibfield  {journal}
  {\bibinfo  {journal} {{Phys. Rev. Lett.}}\ }\textbf {\bibinfo {volume}
  {130}},\ \bibinfo {pages} {061401} (\bibinfo {year} {2023})},\ \Eprint
  {https://arxiv.org/abs/2207.13713} {arXiv:2207.13713 [gr-qc]} \BibitemShut
  {NoStop}%
\bibitem [{\citenamefont {Keir}(2016)}]{Keir2016}%
  \BibitemOpen
  \bibfield  {author} {\bibinfo {author} {\bibfnamefont {J.}~\bibnamefont
  {Keir}},\ }\href {https://doi.org/10.1088/0264-9381/33/13/135009} {\bibfield
  {journal} {\bibinfo  {journal} {Classical Quantum Gravity}\ }\textbf
  {\bibinfo {volume} {33}},\ \bibinfo {pages} {135009} (\bibinfo {year}
  {2016})},\ \Eprint {https://arxiv.org/abs/1404.7036} {arXiv:1404.7036
  [gr-qc]} \BibitemShut {NoStop}%
\bibitem [{\citenamefont {Cardoso}\ \emph {et~al.}(2016)\citenamefont
  {Cardoso}, \citenamefont {Franzin},\ and\ \citenamefont
  {Pani}}]{CardosoFranzinPani2016}%
  \BibitemOpen
  \bibfield  {author} {\bibinfo {author} {\bibfnamefont {V.}~\bibnamefont
  {Cardoso}}, \bibinfo {author} {\bibfnamefont {E.}~\bibnamefont {Franzin}},\
  and\ \bibinfo {author} {\bibfnamefont {P.}~\bibnamefont {Pani}},\ }\href
  {https://doi.org/10.1103/PhysRevLett.116.171101} {\bibfield  {journal}
  {\bibinfo  {journal} {Phys. Rev. Lett.}\ }\textbf {\bibinfo {volume} {116}},\
  \bibinfo {pages} {171101} (\bibinfo {year} {2016})},\ \Eprint
  {https://arxiv.org/abs/1602.07309} {arXiv:1602.07309 [gr-qc]} \BibitemShut
  {NoStop}%
\bibitem [{\citenamefont {Sengo}\ \emph {et~al.}(2023)\citenamefont {Sengo},
  \citenamefont {Cunha}, \citenamefont {Herdeiro},\ and\ \citenamefont
  {Radu}}]{SengoEtAl2023}%
  \BibitemOpen
  \bibfield  {author} {\bibinfo {author} {\bibfnamefont {I.}~\bibnamefont
  {Sengo}}, \bibinfo {author} {\bibfnamefont {P.~V.~P.}\ \bibnamefont {Cunha}},
  \bibinfo {author} {\bibfnamefont {C.~A.~R.}\ \bibnamefont {Herdeiro}},\ and\
  \bibinfo {author} {\bibfnamefont {E.}~\bibnamefont {Radu}},\ }\href
  {https://doi.org/10.1088/1475-7516/2023/01/047} {\bibfield  {journal}
  {\bibinfo  {journal} {{J. Cosmol. Astropart. Phys.}}\ }\textbf {\bibinfo
  {volume} {2023}},\ \bibinfo {pages} {047} (\bibinfo {year} {2023})},\ \Eprint
  {https://arxiv.org/abs/2209.06237} {arXiv:2209.06237 [gr-qc]} \BibitemShut
  {NoStop}%
\bibitem [{\citenamefont {Stuchl\'{i}k}\ and\ \citenamefont
  {Hled\'{i}k}(2002)}]{StuchlikHledik2002}%
  \BibitemOpen
  \bibfield  {author} {\bibinfo {author} {\bibfnamefont {Z.}~\bibnamefont
  {Stuchl\'{i}k}}\ and\ \bibinfo {author} {\bibfnamefont {S.}~\bibnamefont
  {Hled\'{i}k}},\ }\href@noop {} {\bibfield  {journal} {\bibinfo  {journal}
  {{Acta Phys. Slov.}}\ }\textbf {\bibinfo {volume} {52}} (\bibinfo {year}
  {2002})},\ \Eprint {https://arxiv.org/abs/0803.2685} {arXiv:0803.2685
  [gr-qc]} \BibitemShut {NoStop}%
\bibitem [{\citenamefont {Stuchl\'{i}k}\ and\ \citenamefont
  {Hled\'{i}k}(2000)}]{StuchlikHledik2000}%
  \BibitemOpen
  \bibfield  {author} {\bibinfo {author} {\bibfnamefont {Z.}~\bibnamefont
  {Stuchl\'{i}k}}\ and\ \bibinfo {author} {\bibfnamefont {S.}~\bibnamefont
  {Hled\'{i}k}},\ }\href {https://doi.org/10.1088/0264-9381/17/21/312}
  {\bibfield  {journal} {\bibinfo  {journal} {Classical Quantum Gravity}\
  }\textbf {\bibinfo {volume} {17}},\ \bibinfo {pages} {4541} (\bibinfo {year}
  {2000})},\ \Eprint {https://arxiv.org/abs/0803.2539} {arXiv:0803.2539
  [gr-qc]} \BibitemShut {NoStop}%
\bibitem [{\citenamefont {Tang}\ \emph {et~al.}(2017)\citenamefont {Tang},
  \citenamefont {Ong},\ and\ \citenamefont {Wang}}]{TangOngWang2017}%
  \BibitemOpen
  \bibfield  {author} {\bibinfo {author} {\bibfnamefont {Z.-Y.}\ \bibnamefont
  {Tang}}, \bibinfo {author} {\bibfnamefont {Y.~C.}\ \bibnamefont {Ong}},\ and\
  \bibinfo {author} {\bibfnamefont {B.}~\bibnamefont {Wang}},\ }\href
  {https://doi.org/10.1088/1361-6382/aa95ff} {\bibfield  {journal} {\bibinfo
  {journal} {Classical Quantum Gravity}\ }\textbf {\bibinfo {volume} {34}},\
  \bibinfo {pages} {245006} (\bibinfo {year} {2017})},\ \Eprint
  {https://arxiv.org/abs/1705.09633} {arXiv:1705.09633 [gr-qc]} \BibitemShut
  {NoStop}%
\bibitem [{\citenamefont {Astorino}(2012)}]{Astorino2012}%
  \BibitemOpen
  \bibfield  {author} {\bibinfo {author} {\bibfnamefont {M.}~\bibnamefont
  {Astorino}},\ }\href {https://doi.org/10.1007/JHEP06(2012)086} {\bibfield
  {journal} {\bibinfo  {journal} {{J. High Energy Phys.}}\ }\textbf {\bibinfo
  {volume} {2012}},\ \bibinfo {pages} {086} (\bibinfo {year} {2012})},\ \Eprint
  {https://arxiv.org/abs/1205.6998} {arXiv:1205.6998 [gr-qc]} \BibitemShut
  {NoStop}%
\bibitem [{\citenamefont {Ernst}(1968{\natexlab{a}})}]{Ernst1968a}%
  \BibitemOpen
  \bibfield  {author} {\bibinfo {author} {\bibfnamefont {F.~J.}\ \bibnamefont
  {Ernst}},\ }\href {https://doi.org/10.1103/PhysRev.167.1175} {\bibfield
  {journal} {\bibinfo  {journal} {{Phys. Rev.}}\ }\textbf {\bibinfo {volume}
  {167}},\ \bibinfo {pages} {1175} (\bibinfo {year}
  {1968}{\natexlab{a}})}\BibitemShut {NoStop}%
\bibitem [{\citenamefont {Ernst}(1968{\natexlab{b}})}]{Ernst1968b}%
  \BibitemOpen
  \bibfield  {author} {\bibinfo {author} {\bibfnamefont {F.~J.}\ \bibnamefont
  {Ernst}},\ }\href {https://doi.org/10.1103/PhysRev.168.1415} {\bibfield
  {journal} {\bibinfo  {journal} {{Phys. Rev.}}\ }\textbf {\bibinfo {volume}
  {168}},\ \bibinfo {pages} {1415} (\bibinfo {year}
  {1968}{\natexlab{b}})}\BibitemShut {NoStop}%
\bibitem [{\citenamefont {Suvorov}\ and\ \citenamefont
  {Melatos}(2016)}]{SuvorovMelatos2016}%
  \BibitemOpen
  \bibfield  {author} {\bibinfo {author} {\bibfnamefont {A.~G.}\ \bibnamefont
  {Suvorov}}\ and\ \bibinfo {author} {\bibfnamefont {A.}~\bibnamefont
  {Melatos}},\ }\href {https://doi.org/10.1103/PhysRevD.94.044045} {\bibfield
  {journal} {\bibinfo  {journal} {{Phys. Rev. D}}\ }\textbf {\bibinfo {volume}
  {94}},\ \bibinfo {pages} {044045} (\bibinfo {year} {2016})},\ \Eprint
  {https://arxiv.org/abs/1608.03021} {arXiv:1608.03021 [gr-qc]} \BibitemShut
  {NoStop}%
\bibitem [{\citenamefont {Cunha}\ \emph
  {et~al.}(2017{\natexlab{b}})\citenamefont {Cunha}, \citenamefont {Herdeiro},\
  and\ \citenamefont {Radu}}]{CunhaHerdeiroRadu2017}%
  \BibitemOpen
  \bibfield  {author} {\bibinfo {author} {\bibfnamefont {P.~V.~P.}\
  \bibnamefont {Cunha}}, \bibinfo {author} {\bibfnamefont {C.~A.~R.}\
  \bibnamefont {Herdeiro}},\ and\ \bibinfo {author} {\bibfnamefont
  {E.}~\bibnamefont {Radu}},\ }\href
  {https://doi.org/10.1103/PhysRevD.96.024039} {\bibfield  {journal} {\bibinfo
  {journal} {{Phys. Rev. D}}\ }\textbf {\bibinfo {volume} {96}},\ \bibinfo
  {pages} {024039} (\bibinfo {year} {2017}{\natexlab{b}})},\ \Eprint
  {https://arxiv.org/abs/1705.05461} {arXiv:1705.05461 [gr-qc]} \BibitemShut
  {NoStop}%
\bibitem [{\citenamefont {Shipley}\ and\ \citenamefont
  {Dolan}(2016)}]{ShipleyDolan2016}%
  \BibitemOpen
  \bibfield  {author} {\bibinfo {author} {\bibfnamefont {J.~O.}\ \bibnamefont
  {Shipley}}\ and\ \bibinfo {author} {\bibfnamefont {S.~R.}\ \bibnamefont
  {Dolan}},\ }\href {https://doi.org/10.1088/0264-9381/33/17/175001} {\bibfield
   {journal} {\bibinfo  {journal} {{Classical Quantum Gravity}}\ }\textbf
  {\bibinfo {volume} {33}},\ \bibinfo {pages} {175001} (\bibinfo {year}
  {2016})},\ \Eprint {https://arxiv.org/abs/1603.04469} {arXiv:1603.04469
  [gr-qc]} \BibitemShut {NoStop}%
\bibitem [{\citenamefont {Lima~Junior}\ \emph {et~al.}(2021)\citenamefont
  {Lima~Junior}, \citenamefont {Cunha}, \citenamefont {Herdeiro},\ and\
  \citenamefont {Crispino}}]{LimaJuniorEtAl2021}%
  \BibitemOpen
  \bibfield  {author} {\bibinfo {author} {\bibfnamefont {H.~C.~D.}\
  \bibnamefont {Lima~Junior}}, \bibinfo {author} {\bibfnamefont {P.~V.~P.}\
  \bibnamefont {Cunha}}, \bibinfo {author} {\bibfnamefont {C.~A.~R.}\
  \bibnamefont {Herdeiro}},\ and\ \bibinfo {author} {\bibfnamefont {L.~C.~B.}\
  \bibnamefont {Crispino}},\ }\href
  {https://doi.org/10.1103/PhysRevD.104.044018} {\bibfield  {journal} {\bibinfo
   {journal} {{Phys. Rev. D}}\ }\textbf {\bibinfo {volume} {104}},\ \bibinfo
  {pages} {044018} (\bibinfo {year} {2021})},\ \Eprint
  {https://arxiv.org/abs/2104.09577} {arXiv:2104.09577 [gr-qc]} \BibitemShut
  {NoStop}%
\bibitem [{\citenamefont {Grenzebach}\ \emph {et~al.}(2014)\citenamefont
  {Grenzebach}, \citenamefont {Perlick},\ and\ \citenamefont
  {L\"{a}mmerzahl}}]{GrenzebachPerlickLammerzahl2014}%
  \BibitemOpen
  \bibfield  {author} {\bibinfo {author} {\bibfnamefont {A.}~\bibnamefont
  {Grenzebach}}, \bibinfo {author} {\bibfnamefont {V.}~\bibnamefont
  {Perlick}},\ and\ \bibinfo {author} {\bibfnamefont {C.}~\bibnamefont
  {L\"{a}mmerzahl}},\ }\href {https://doi.org/10.1103/PhysRevD.89.124004}
  {\bibfield  {journal} {\bibinfo  {journal} {{Phys. Rev. D}}\ }\textbf
  {\bibinfo {volume} {89}},\ \bibinfo {pages} {124004} (\bibinfo {year}
  {2014})},\ \Eprint {https://arxiv.org/abs/1403.5234} {arXiv:1403.5234
  [gr-qc]} \BibitemShut {NoStop}%
\bibitem [{\citenamefont {Charmousis}\ \emph {et~al.}(2007)\citenamefont
  {Charmousis}, \citenamefont {Langlois}, \citenamefont {Steer},\ and\
  \citenamefont {Zegers}}]{CharmousisEtAl2007}%
  \BibitemOpen
  \bibfield  {author} {\bibinfo {author} {\bibfnamefont {C.}~\bibnamefont
  {Charmousis}}, \bibinfo {author} {\bibfnamefont {D.}~\bibnamefont
  {Langlois}}, \bibinfo {author} {\bibfnamefont {D.}~\bibnamefont {Steer}},\
  and\ \bibinfo {author} {\bibfnamefont {R.}~\bibnamefont {Zegers}},\ }\href
  {https://doi.org/10.1088/1126-6708/2007/02/064} {\bibfield  {journal}
  {\bibinfo  {journal} {{J. High Energy Phys.}}\ }\textbf {\bibinfo {volume}
  {2007}},\ \bibinfo {pages} {064} (\bibinfo {year} {2007})},\ \Eprint
  {https://arxiv.org/abs/gr-qc/0610091} {arXiv:gr-qc/0610091} \BibitemShut
  {NoStop}%
\end{thebibliography}%

\end{document}